# Scalable, Time-Responsive, Digital, Energy-Efficient Molecular Circuits using DNA Strand Displacement[*]

Ehsan Chiniforooshan[†]   David Doty[†]   Lila Kari[†]   Shinnosuke Seki[†]

**Abstract**

We propose a novel theoretical biomolecular design to implement any Boolean circuit using the mechanism of DNA strand displacement. The design is *scalable*: all species of DNA strands can in principle be mixed and prepared in a single test tube, rather than requiring separate purification of each species, which is a barrier to large-scale synthesis. The design is *time-responsive*: the concentration of output species changes in response to the concentration of input species, so that time-varying inputs may be continuously processed. The design is *digital*: Boolean values of wires in the circuit are represented as high or low concentrations of certain species, and we show how to construct a single-input, single-output signal restoration gate that amplifies the difference between high and low, which can be distributed to each wire in the circuit to overcome signal degradation. This means we can achieve a digital abstraction of the analog values of concentrations. Finally, the design is *energy-efficient*: if input species are specified ideally (meaning absolutely 0 concentration of unwanted species), then output species converge to their ideal concentrations at steady-state, and the system at steady-state is in (dynamic) equilibrium, meaning that no energy is consumed by irreversible reactions until the input again changes.

Drawbacks of our design include the following. If input is provided non-ideally (small positive concentration of unwanted species), then energy must be continually expended to maintain correct output concentrations even at steady-state. In addition, our fuel species – those species that are permanently consumed in irreversible reactions – are not "generic"; each gate in the circuit is powered by its own specific type of fuel species. Hence different circuits must be powered by different types of fuel. Finally, we require input to be given according to the *dual-rail* convention, so that an input of 0 is specified not only by the absence of a certain species, but by the presence of another. That is, we do not construct a "true NOT gate" that sets its output to high concentration if and only if its input's concentration is low. It remains an open problem to design scalable, time-responsive, digital, energy-efficient molecular circuits that additionally solve one of these problems, or to prove that some subset of their resolutions are mutually incompatible.

## 1 Introduction

Biomolecular circuits, due to natural compatibility with the materials of life, could change the way diseases are diagnosed or medicine is delivered, through bottom-up, site-specific biochemical information processing. While there is no shortage of theoretical proposals and experimental implementations [4–10, 13, 15], the state of the art in biomolecular circuits remains far behind its electronic equivalent.

---
[*]This research was supported in part by Natural Sciences and Engineering Research Council of Canada (NSERC) Discovery Grant R2824A01 and the Canada Research Chair Award in Biocomputing to Lila Kari.

[†]University of Western Ontario, Dept. of Computer Science, London, Ontario, Canada, N6A 5B7, {ehsan,ddoty,lila,sseki}@csd.uwo.ca.

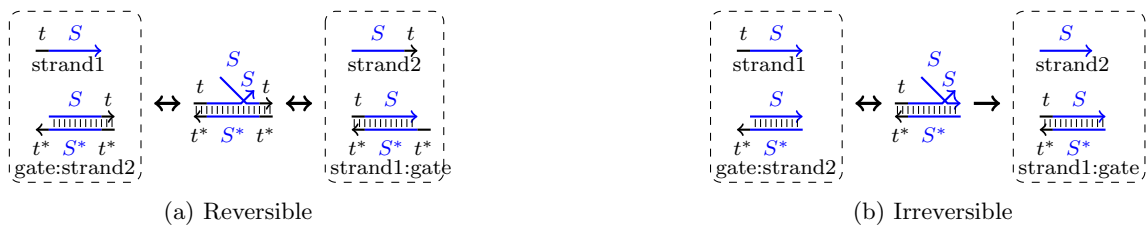

(a) Reversible  (b) Irreversible

Figure 1: Example of basic DNA strand displacement reactions. Figure 1a is the reversible reaction strand1+gate:strand2 ⇌ strand1:gate+strand2, and Figure 1b is the irreversible reaction strand1+ gate:strand2 → strand1:gate + strand2. Each strand is represented as an arrow indicating 5'-to-3' orientation. $t$ and $S_1$ are finite strings over $\{A, C, G, T\}$, and $t^*$ and $S^*$ are their 3'-to-5'-oriented Watson-Crick complements. The "toehold" region $t$ is short enough (about 5 nucleotides) that it cannot provide sufficient binding strength to allow two strands to hybridize stably. Only if the longer (say, at least 15 nucleotides) "recognition" region $S$ matches between strand1 and the base strand gate can strand1 displace strand2. Displacement occurs via an unbiased random walk but proceeds quickly compared to the time taken for strands to find each other in solution, so a successful displacement is modeled as occurring instantaneously upon binding of the toehold.

In this paper, we propose a new theoretical design, focusing on one particular molecular primitive as our sole "basic operation" from which to compose complex circuits: *toehold-mediated DNA branch migration and strand displacement*, or simply *strand displacement*. The idea of the basic strand displacement reaction is shown in Figure 1. DNA strand displacement has been useful for powering DNA-based nanodevices (e.g., [14]), and its use as a primitive for building circuits was pioneered by Seelig, Soloveichik, Zhang, and Winfree [10], and subsequently improved and simplified by Zhang, Turberfield, Yurke, and Winfree [15] and Qian and Winfree [9]. Compared to some designs using other "molecular primitives" such as restriction enzymes, strand displacement has the advantage that it requires the design of no new molecules other than single-stranded DNA complexes, which are easily and cheaply available through mail-order, e.g. [3].

Our construction achieves four properties desirable of robust circuit designs. We do not claim these properties to represent the sole criteria by which to judge molecular circuit designs, but we believe they are evidently advantageous. We emphasize that achieving any of these in isolation would not be a novel contribution, but to our knowledge this is the first DNA circuit design to achieve all four simultaneously. The properties are as follows.

**scalable:** This is a nebulous word with many connotations. Our particular usage refers to the definition given by Qian and Winfree in [9], whose construction overcomes a specific inhibiting factor that prevents large-scale fabrication of molecular circuits: the need to prepare and purify different molecular species in separate test tubes. Since a Boolean circuit with hundreds of gates may require thousands of distinct molecular species, it is no small advantage to be able to mix all of them together in a single tube and conduct the necessary preparation steps solely on that one tube; this is what we mean by "scalable".[1] The particular mechanism

---

[1]Although we have not formally defined the notion, it seems reasonable, for instance, to define a "scalable"

we utilize to achieve scalability is the same used by Qian and Winfree, and is described in Section 2.3. Briefly, it involves creating double-stranded complexes from single-stranded hairpin precursors that are cleaved with (naturally occurring) restriction enzymes, possibly leaving short sticky ends to serve as toeholds for future strand displacement.

**time-responsive:** This property is achieved by Goel and Ibrahimi [4] utilizing restriction enzyme technology (apparently requiring new restriction enzymes to be designed). Qian and Winfree [9] describe time-responsiveness as an open problem for their particular motif, known as seesaw gates, which are built from strand displacement cascades. Informally, a circuit is time-responsive if, supposing that the inputs to the circuit change after the initial computation, then the output is re-computed to reflect the new inputs. Time-responsiveness of individual gates is key to constructing recurrent circuits that use feedback loops to implement memory storage devices, such as latches and flip-flops. Even for feed-forward circuits, time-responsiveness is an intuitively appealing property. For instance, a circuit could constantly monitor the state of a cell and release drugs in response to a temporary malady, then inhibit the release as the malady disappears.

**energy-efficient:** This property is a definition of our own device, but it or something approximating it seems essential in robust circuit implementations. Our definition is that inputs given ideally result in eventual migration to a steady state in which 1) outputs are also ideal, and 2) this steady state is in equilibrium, maintained without any expenditure of energy to power irreversible reactions. This is a dynamic equilibrium, as some reversible reactions are always taking place. In our system (as in many others), Boolean values of all wires in the circuit, including input, output, and intermediate gate-connecting wires, are represented as high or low concentrations of certain chemical species. As each wire $w$ in the abstract circuit being simulated can take on the values 0 and 1, this wire is associated to two chemical species $0_w$ and $1_w$. Ideally, to represent the bit $b$, $b_w$ is present and $\overline{b}_w$ is absent (where $\overline{b} = 1 - b$), the so-called *dual-rail* convention. Our design ensures energy efficiency because energy is expended only to change non-ideal wires (those with positive concentration of $b_w$ when $\overline{b}$ is the correct bit for that wire given the inputs) to ideal, but no energy is expended to maintain a wire's correctness once it reaches an ideal state. Our definition of energy efficiency makes the helpful but far-reaching assumption of ideally-presented inputs; Section 3.1 discusses the effect of non-ideal inputs on the behavior of our circuit.

**digital:** By this we mean that the circuit employs signal restoration to obtain a digital abstraction of what are fundamentally analog concentration values. Since Boolean values are represented by high or low concentrations of certain chemical species, to correct for non-ideal inputs, as well as the natural signal degradation suffered by the logic gates, it is desirable to move high concentrations higher and low concentrations lower before feeding the values as input to the next gate in the circuit. We achieve this by designing a single-input, single-output restoration gate (to be "spliced" into every wire in the logical circuit) such that, if $i$ is the input wire and $o$ is the output wire, then $[1_o]/[0_o] = ([1_i]/[0_i])^2$. For instance, if a wire's species have combined concentration 100, and if $[1_i] = 60$ and $[0_i] = 40$, then at steady-state, $[1_o] \approx 69.23$ and $[0_o] \approx 30.77$ (since $69.23/30.77 \approx 2.25 = 1.5^2 = (60/40)^2$). By serially cascading a small

---

molecular circuit to be one whose preparation involves at most a constant number of preparation steps (other than the original design of the molecules themselves), regardless of the size of the circuit.

number of such gates together we can amplify even very weak signals, since $n$ cascaded gates amplify a ratio of $r$ to $r^{2^n}$. For instance, to transform a ratio of $0.6/0.4$ to $> 0.99999/0.00001$ requires only 5 restoration gates.

Our design has limitations as well, discussed in greater detail in Section 3.1. In Section 3.2, some designs that are similar in goals or methods [4, 7, 9, 11, 12] are discussed and compared with the present design.

We omit a rigorous analysis of the soundness of our design in this extended abstract. However, our design has been simulated using the model of DNA strand displacement kinetics described in [9, 11, 12], and the simulation agrees with the properties claimed above. The results are shown in Section 4. An important future project is to experimentally validate that such a construction is possible.

## 2 Construction

This section describes the details of our proposed design. Recall the dual-rail convention that our design employs: each wire $w$ in the circuit will be represented by two chemical species (DNA strands) $0_w$ and $1_w$, and the Boolean gates we use work as long as the ratio of the concentration of the correct input bit to the incorrect input bit is "sufficiently high". Our design involves the construction of three types of gates, the first two Boolean and the third analog: 2-input/1-output Boolean NAND gates (output bit is 0 if and only if both input bits are 1), 1-input/2-output Boolean fan-out gates (both output bits are equal to input bit), and 1-input/1-output analog signal restoration gates (the difference in concentration between species $0_w$ and $1_w$ representing the input wire $w$ is amplified for the output wire species). Every Boolean function can be computed by the composition of NAND gates, and through the appropriate composition of fan-out gates we may assume the fan-out of every NAND gate and every input is 1. We then place a signal restoration gate (or more in series) on each wire of the circuit. This is necessary since the Boolean gates suffer some signal degradation, and since inputs may not be specified ideally. The conversion of an example circuit is shown in Figure 2. We choose NAND gates for the sake of concreteness, but it is easy to modify our design to implement any of the sixteen possible 2-input/1-output logic gates.

Our design of the three gates is described in terms of sets of abstract chemical reactions implemented by DNA strand displacement in a similar fashion to [11, 12]. Each species in the reactions below is represented by single-stranded DNA molecules of the same form as strand1 in Figure 1;[2] since they all have the same format, arbitrarily large circuits may be "wired" together. Our method of implementing abstract chemical reactions with DNA strand displacement reactions is inspired by, but subtly different from, that of [11, 12]. The differences and reasons for these differences are outlined in more detail in Section 3.2. As in [11, 12], a single "high-level" abstract reaction is simulated by more than one "underlying" DNA strand displacement reaction. The primary difference is the need for a direct implementation of a termolecular reaction (a reaction with three reactants) in which no irreversible "underlying implementation reaction" is allowed to occur unless all three reactants of the "high-level reaction" are present. This allows us to conclude that if no abstract high-level reactions are possible, then no irreversible, energy-consuming implementation reactions are possible. This ensures energy efficiency so long as we design the high-level reactions so that they cannot occur at steady-state.

---

[2] Other intermediate species not shown in the section but used in the DNA strand implementation in Section 2.2 are not necessarily of this form.

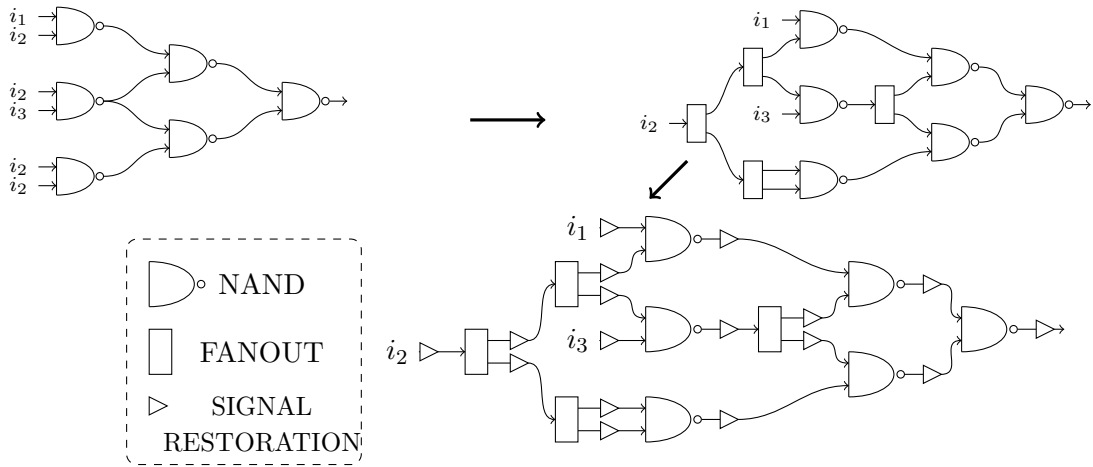

Figure 2: A circuit with 6 Boolean NAND gates and 3 inputs, converted to add Boolean fan-out gates to allow fan-out from inputs and NAND gates, as well as analog signal restoration gates between all Boolean gates.

We do not specify the rate constant of any reaction. For the low-level strand displacement reactions all rate constants are assumed to be equal, following the kinetic model of [11, 12] in which toehold length determines the rate constant of a strand displacement reaction (all our toeholds are equal length). In implementation using DNA strands, the abstract high-level reactions will not have identical rate constants, but the implementation is robust to small differences in rate constants among the high-level reactions. For instance, it suffices if all of these rate constants are within an order of magnitude of each other, which they are given our implementation.

The time-responsiveness of the design is evident by inspection of the reactions below. Informally, it follows from the fact that each input to a gate is catalytic in the reactions associated with that gate and the fact that the total number of "bit molecules" $0_w$ and $1_w$ associated with a wire $w$ is constant; reactions only change $0_w$ to $1_w$ and vice-versa. Fuel molecules are indeed consumed for each reaction. Some consumed fuel species are explicit in the reactions discussed below (such as $\mathsf{diff}_o$ or $\mathsf{0f}_o$), but some fuel is explicit only in the "underlying implementation reactions" of DNA strand displacement and does not even appear as an abstract chemical species below. For each set of reactions below the fuel efficiency of the reactions is justified in the sense that none of the reactions can occur at the steady-state concentrations assuming ideal input species. We must take care when implementing these reactions with DNA strand displacement to ensure that all underlying implementation reactions that occur at steady-state are reversible, but this is not discussed in this section. The digital abstraction is achieved by one particular signal restoration gate discussed in Section 2.1.3, intended to be distributed throughout the circuit in between the other gates. The scalability of the design is a consequence of how we propose to construct the DNA complexes required for implementation and is discussed in Section 2.3.

## 2.1 Design of Abstract Chemical Reactions for Gates

Given the circuit structure proposed above (NAND gates, fan-out gates, and signal restoration gates), we may assume that each wire in the circuit is the input of exactly one gate and the output of exactly one gate. If each gate $g$ in the circuit is given a unique identifier $\mathsf{id}_g$, then a wire connecting gate $g$ to $g'$ is uniquely identified by the pair $w = (\mathsf{id}_g, \mathsf{id}_{g'})$. We therefore assume each wire has a unique identifier that indicates which two gates it connects. Ultimately, this wiring information will be encoded by DNA recognition regions, but for now we describe the gate operation in terms of wires labeled with abstract values. The concentrations of the species associated with the input wires are assumed to be under external control. We take care to ensure that all wire species are catalytic (not consumed) in the reactions of gates for which they are an *input*; this ensures time-responsiveness of each of the gates in the presence of sufficient fuel.

### 2.1.1 NAND Gate Reactions

Given a wire identifier $w$, associate with it the two chemical species $\mathsf{0}_w$ and $\mathsf{1}_w$, representing the value of the wire through the dual-rail convention described earlier. To implement a NAND gate, whose output is 0 if and only if both inputs are 1, with input wires $i_1$ and $i_2$ and output wire $o$, we use the reactions

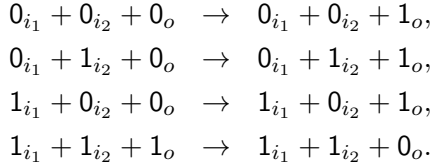

$$\begin{aligned}
\mathsf{0}_{i_1} + \mathsf{0}_{i_2} + \mathsf{0}_o &\to \mathsf{0}_{i_1} + \mathsf{0}_{i_2} + \mathsf{1}_o, \\
\mathsf{0}_{i_1} + \mathsf{1}_{i_2} + \mathsf{0}_o &\to \mathsf{0}_{i_1} + \mathsf{1}_{i_2} + \mathsf{1}_o, \\
\mathsf{1}_{i_1} + \mathsf{0}_{i_2} + \mathsf{0}_o &\to \mathsf{1}_{i_1} + \mathsf{0}_{i_2} + \mathsf{1}_o, \\
\mathsf{1}_{i_1} + \mathsf{1}_{i_2} + \mathsf{1}_o &\to \mathsf{1}_{i_1} + \mathsf{1}_{i_2} + \mathsf{0}_o.
\end{aligned}$$

In other words, if two inputs "encounter" an output that is erroneous (according to the two inputs), then two inputs cooperate to "fix" the output molecule by converting it to represent the other bit. It is clear that with ideal inputs (i.e., $[\mathsf{0}_i] > 0 \iff [\mathsf{1}_i] = 0$ for $i \in \{i_1, i_2\}$), then at steady state, the output is ideal, correct, and none of the reactions above are possible since all would have at least one reactant with concentration 0. Therefore these reactions are energy-efficient.

By modifying the bits of the rightmost reactant and product in each reaction above to represent a different truth table, the reactions can be made to emulate any 2-input/1-output logic gate.

### 2.1.2 Fan-out Gate Reactions

A fan-out gate takes a single input wire $i$ and copies its value to two output wires $o_1$ and $o_2$. The reactions to implement this are

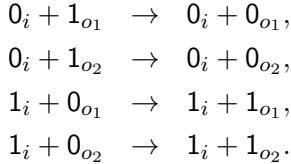

$$\begin{aligned}
\mathsf{0}_i + \mathsf{1}_{o_1} &\to \mathsf{0}_i + \mathsf{0}_{o_1}, \\
\mathsf{0}_i + \mathsf{1}_{o_2} &\to \mathsf{0}_i + \mathsf{0}_{o_2}, \\
\mathsf{1}_i + \mathsf{0}_{o_1} &\to \mathsf{1}_i + \mathsf{1}_{o_1}, \\
\mathsf{1}_i + \mathsf{0}_{o_2} &\to \mathsf{1}_i + \mathsf{1}_{o_2}.
\end{aligned}$$

As with the NAND gate reactions, it is clear that the reactions are energy-efficient since at steady-state at least one reactant of each reaction has concentration 0.

### 2.1.3 Signal Restoration Gate Reactions

The NAND and fan-out gates contain no digital abstraction: each input species "pushes" on the output in linear proportion to the concentration of the input. This implies that at best the signal would not be lost; i.e., if input were 90% ideal (e.g. $[0_i] = 0.9$ and $[1_i] = 0.1$), then the best the output could be is 90% ideal. In reality, even this is not achieved since our DNA strand displacement implementation of the reactions introduces some non-idealities that imply there will be signal loss at each logical gate. The signal restoration gate described below functions to restore the signal. The gate takes one input $i$ and produces one output $o$, such that at steady-state, $[1_o]/[0_o] = ([1_i]/[0_i])^2$. Therefore the difference between the input species that is "high" and the one that is "low" will be amplified.[3]

The following reactions implement this gate. Below we describe the intuition behind them.

$$0_i + 1_o \rightarrow 0_i + 1_o + \mathsf{diff}_o, \qquad (2.1)$$
$$0_i + \mathsf{diff}_o \rightarrow 0_i + \mathsf{p0}_o, \qquad (2.2)$$
$$\mathsf{p0}_o + \mathsf{p0}_o \rightarrow \mathsf{p0}_o + \mathsf{p0}_o + \mathsf{P0}_o, \qquad (2.3)$$
$$\mathsf{P0}_o + 1_o \rightarrow \mathsf{P0}_o + 0_o, \qquad (2.4)$$
$$1_i + 0_o \rightarrow 1_i + 0_o + \mathsf{diff}_o, \qquad (2.5)$$
$$1_i + \mathsf{diff}_o \rightarrow 1_i + \mathsf{p1}_o, \qquad (2.6)$$
$$\mathsf{p1}_o + \mathsf{p1}_o \rightarrow \mathsf{p1}_o + \mathsf{p1}_o + \mathsf{P1}_o, \qquad (2.7)$$
$$\mathsf{P1}_o + 0_o \rightarrow \mathsf{P1}_o + 1_o. \qquad (2.8)$$

Additionally, there is a species $\mathsf{decay}_o$ set to some constant concentration (comparable to that of $0_o$ and $1_o$), such that, for each species $S \in \{\mathsf{diff}_o, \mathsf{p0}_o, \mathsf{p1}_o, \mathsf{P0}_o, \mathsf{P1}_o\}$, we also have the reaction

$$\mathsf{decay}_o + S \rightarrow \mathsf{decay}_o. \qquad (2.9)$$

The intuition behind the reactions above is as follows. Reactions (2.1) through (2.4) are functionally the same as (2.5) through (2.8); the difference is only in which input bit is being translated to the output. For concreteness we describe only reactions (2.1) through (2.4); therefore we regard the input bit as 0, $0_i$ as the "correct" input species, and $1_i$ as the "incorrect" input species. Reaction (2.1) is designed to detect the presence of incorrect output; if an input molecule encounters an output molecule that it "thinks" is incorrect, those molecules catalytically produce a copy of $\mathsf{diff}_o$. The purpose of $\mathsf{diff}_o$ is to "announce" to both input species that the output is not ideal; when an input molecule $0_i$ reacts with $\mathsf{diff}_o$ in reaction (2.2), the input catalytically transforms $\mathsf{diff}_o$ into a "push" molecule $\mathsf{p0}_o$.[4] If this molecule $\mathsf{p0}_o$ were to react directly with the output $1_o$ to convert it to $0_o$, the rate of conversion would be linear in the input concentration; hence signal restoration

---

[3]The statement "$[1_o]/[0_o] = ([1_i]/[0_i])^2$ at steady-state" applies to the abstract high-level reactions described in this section but is not entirely accurate for the implementation reactions of Section 2.2 due to the "buffering" effect described in that section. DNA strands spend part of their time "buffered" in double-stranded complexes through reversible exchange reactions even in the absence of the other reactants. The effect is that their "net concentration" may be lower at steady-state by some constant multiplicative factor than their initial concentration (the purpose of the fan-out gates is to keep this factor constant). Nonetheless, we still obtain a quadratic amplification of the ratio $[1_i]/[0_i]$ although perhaps off by a constant to account for the buffering effect.

[4]Note that reactions (2.1) and (2.5) produce the same molecule $\mathsf{diff}_o$; this is so that, if the output is "close to ideal", the production of output-changing molecules will not become unbalanced in favor of the incorrect output. If

would not occur since the reverse conversion of $0_o$ to $1_o$ catalyzed by $\mathsf{p1}_o$ would proceed at a rate proportional to the concentration of $1_i$. To amplify the ratio between the correct input species and the incorrect input species, we must create push molecules whose ratio of correct-to-incorrect is superlinear in the correct-to-incorrect ratio of the input species. This is the function of reaction (2.3), which produces "strong push" molecules $\mathsf{P0}_o$ at a rate quadratic in the concentration $\mathsf{p0}_o$ (since the rate of a bimolecular reaction of the form $A + A \to \ldots$ is proportional to $[A]^2$). Reaction (2.4) is then the "correction" of the output species by the strong push molecules.

The reactions are energy-efficient. Although our simulation uses the mass-action kinetics model, it is easiest to describe the intuition with finite counts of molecules, so that molecular concentrations can really go from positive to 0 in a finite amount of time. With ideal input species, say, $[0_i] > 0$ and $[1_i] = 0$, reactions (2.5) and (2.6) cannot occur. Since (2.6) cannot occur, production of $\mathsf{p1}_o$ halts and $[\mathsf{p1}_o]$ decays to concentration 0 through reaction (2.9). With $[\mathsf{p1}_o] = 0$, reaction (2.7) cannot occur and production of $\mathsf{P1}_o$ halts and $[\mathsf{P1}_o]$ decays to 0 through reaction (2.9). No reaction at this point can change a $0_o$ to a $1_o$. Reactions (2.1) through (2.4) continue until all $1_o$ are converted to $0_o$. At this point reaction (2.1) cannot occur, production of $\mathsf{diff}_o$ halts, and $[\mathsf{diff}_o]$ decays to concentration 0 through reaction (2.9). This results in the eventual decay of $\mathsf{p0}_o$ and $\mathsf{P0}_o$, for the same reason as described above for $\mathsf{p1}_o$ and $\mathsf{P1}_o$, at which point none of the reactions (2.1) through (2.9) can occur, and the system has reached steady-state.

## 2.2 Implementation of Abstract Reactions with DNA Strand Displacement

Our design of a NAND gate, a fan-out gate, and a signal restoration gate described in Section 2.1 consists of bimolecular and termolecular reactions (meaning two and three reactants, respectively) with abstract chemicals. We now introduce a method to implement these abstract reactions with DNA strand displacement. For concreteness we show how to implement a bimolecular reaction $A + B \to C + D$ with two products and a termolecular reaction $A + B + C \to D + E + F$ with three products. However, it should be clear how to modify each design to allow an arbitrary number of products in either case. For instance, our design from Section 2.1 also requires bimolecular reactions with three products (e.g. $0_i + 1_o \to 0_i + 1_o + \mathsf{diff}_o$). As mentioned earlier, this design borrows heavily from [11, 12] but alters the design for scalability as described in Section 2.3 and to allow direct implementation of termolecular reactions. The latter problem is nontrivial due to the effect of the "buffer" strands for the second reactant, as we will explain below.

The implementation of a bimolecular reaction $A + B \to C + D$ using DNA strand displacement is shown in Figure 3. The series of reactions is activated in the presence of reactants $ta$ ($A$), $tb$ ($B$) and the gate complex $g_1$. First, $ta$ attaches to the gate $g_1$ via the uncovered toehold $t$ and the strand displacement follows to detach the buffer1 $at$ from $g_1$ (reaction (a) in Figure 3). At this stage two reactions are possible: the strand displacement caused by $tb$ to release the linker (reaction (b)) or the buffer1 strand reverses the previous reaction. Reaction (b) is also reversible due to the exposed toehold on the right side of the complex after reaction (b). At this point the linker can bind to the gate $g_2$ and release products $tc$ ($C$) and $td$ ($D$) (reaction (c)). Strand 1 and the two complexes with the bottom strands of gates $g_1$ and $g_2$ are produced as waste. This reaction is irreversible, but since it is the only irreversible reaction in the cascade, either the entire cascade is "committed" or it is possible to return to the original state shown in the upper left box of Figure 3.

---

reactions (2.1) and (2.5) produced molecules specific to one input bit, say $\mathsf{diff0}_o$ and $\mathsf{diff1}_o$, then the production rate of $\mathsf{diff0}_o$ would decrease, and the production rate of $\mathsf{diff1}_o$ would increase, as the output species moved closer to an ideal representation of 0. This would lead to a negative feedback loop hampering signal amplification.

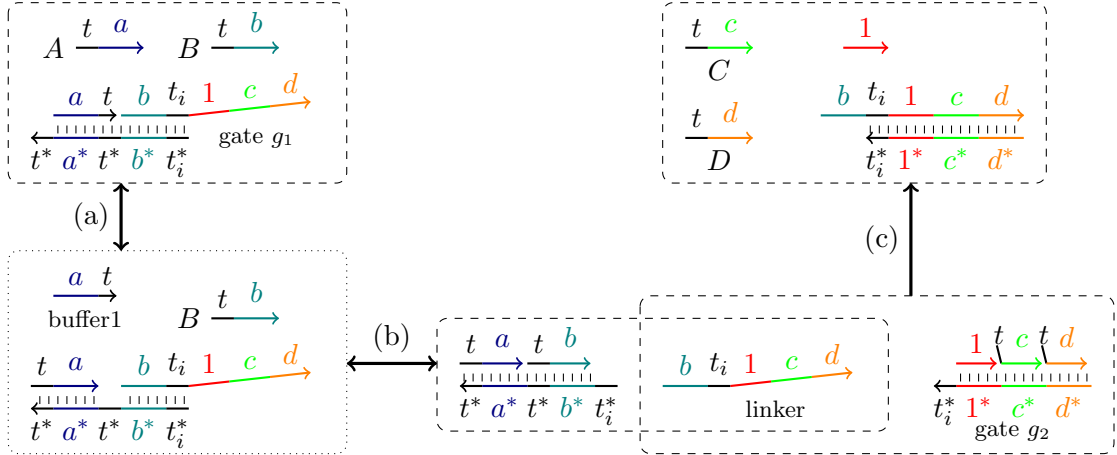

Figure 3: The DNA motif and reaction mechanism for the bimolecular reaction $A+B \to C+D$. By $a, b, c, d, 1$, we denote the recognition domains. $t$ and $t_i$ (for $i \in \{1, 2, 3\}$) are the toehold domains. The need for the different toeholds is explained in Section 2.4.1. The bolded arrow between two dotted squares surrounding respective sets of DNA motifs represents the reaction with the set of DNA motifs at its tail as its reactants and the set at its head as products. Recognition region 1 uniquely identifies this reaction and ensures that the linker strand cannot release any output strands from the gate complex $g_2$ of other reactions even if they share a first output recognition region ($c$ in this case).

In the absence of $B$, some proportion of copies of strand $A$ spend time "buffered" in gate complex $g_1$, reversibly exchanging with buffer1, which lowers the "effective concentration" of $A$, altering the strength of its effect on other reactions of which it is a part. However, given the design of Section 2.1, particularly due to the use of fan-out gates, we may assume each species is a first or second reactant in at most two reactions. Therefore the effective concentration of $A$ is at least a constant fraction of its initial concentration, given a sufficiently large and approximately equal supply of buffer1 and $g_1$ (in each of the two reactions in which $A$ participates). This is essentially the same argument used in [11] (although handled differently in [12] to more precisely simulate desired rate constants, which do not need to be precise for our purposes). As explained below the existence of a second buffer strand in the termolecular reaction requires more sophisticated handling.

The implementation of a termolecular reaction $A + B + C \to D + E + F$ is illustrated in Figure 4. The essential difference is the buffer2 collector. Whenever the termolecular reaction completes successfully, the numbers of buffer1 and buffer2 increase by 1. Unless processed properly these buffers accumulate and become more and more competitive against $A$ and $B$. For the buffer1 strand, this problem may be solved by having the gate $g_1$ and buffer1 be in excess such that their concentrations remain effectively constant; this is the trick used in [11]. However, this argument does not apply to buffer2, since buffer2 reacts with $g_1 : A$ complexes. The number of these is necessarily no larger than the number of $A$ strands, no matter how many gate complexes $g_1$ are supplied. Therefore the produced copies of buffer2 will eventually grow so large as to effectively prevent $C$ from binding unless the copies of buffer2 are collected. Hence a "collector" is prepared for buffer2, which binds to buffer2 to render it inert. It is critical, however, that the collector not be released but upon the successful completion of the entire cascade of reactions, since we require positive concentration of buffer2 in the case that $A$ and $B$ are present but $C$ is absent, to prevent the undesirable situation that all copies of $B$ become trapped in gate $g_1$ complexes. Because of the need to deal with buffer1 strands and buffer2 strands differently, it is essential to prevent crosstalk between them; this is explained in Section 2.4.2.

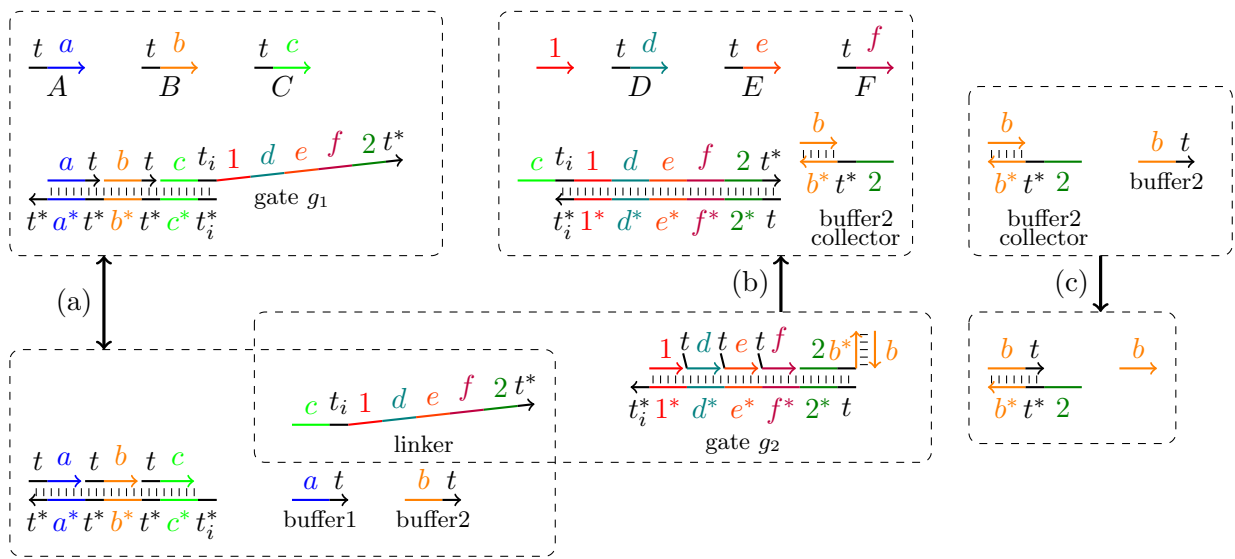

Figure 4: The DNA motif and reaction mechanism for the termolecular reaction $A + B + C \to D + E + F$.

## 2.3 Scalable Preparation of DNA Complexes

The multi-strand gate complexes $g_1$ and $g_2$ in our implementation are prepared in a scalable manner using a single-stranded multiple hairpin precursor, cleaved by restriction enzymes, similar to the design proposed by Qian and Winfree [9]. The fundamental idea is that a single DNA strand, which is designed so as to form a multiple hairpin structure, is more likely to form during annealing from a high temperature than intramolecular structures involving separate strands, even if these multi-strand complexes have lower energy. The single-stranded hairpins become "kinetically trapped" [1, 2]. This prevents the single strands that are intended to eventually displace strands from the complex from binding to the complex initially before computation even begins.

In order to create a gate complex that consists of a bottom strand and three top strands such as the gate complex $g_1$ in Figure 4, we first design a single strand that during annealing folds into itself to form a multiple hairpin precursor as in Figure 5. Stems of these hairpins contain the recognition site of a restriction enzyme so that, after annealing, we may add the enzyme to cut the hairpin and separate it from the precursor. A structure with $n$ hairpins thus processed results in $n$ top strands bound to one bottom strand.

The leftmost hairpin is supposed to be cleaved by an enzyme, and a sticky end is left on the bottom strand, which will work as a toehold. For example, Figure 6 illustrates the hairpin detachment by the enzyme BplI. This enzyme recognizes the double strand whose upper strand is GAGNNNNNCTC, where N denotes an arbitrary nucleotide, and then cuts the upper strand at 13nt to the right of the recognition site and the lower strand at 8nt to the right of the site. A sticky end of length 5nt remains; this serves as the toehold sequence. BplI does not care what sequence of length 5nt is left, but cares only the distance from its recognition site; we have chosen CCTCC arbitrarily as the toehold sequence. This enables us to create a toehold that is not a "Watson-Crick palindrome", a string of the form $t = x(x^*)^{\mathcal{R}}$, where $(x^*)^{\mathcal{R}}$ is the reversed Watson-Crick complement of $x$. Such strings have the undesirable property of being complementary to themselves (under the proper 5'-to-3' orientation for hybridization); i.e. a toehold could bind to another copy of itself.

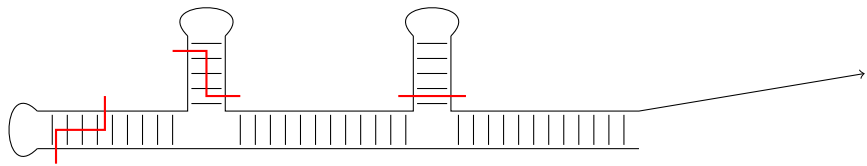

Figure 5: A multiple hairpin precursor for the gate complex design.

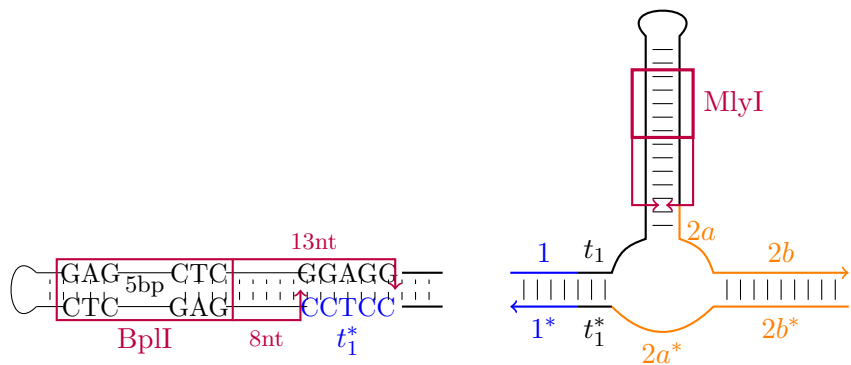

Figure 6: Examples of cut of hairpins on the precursor by restriction enzymes.

The detachment of the leftmost hairpin separates the top strand from the bottom strand of a gate complex. In contrast, detachment of the top hairpins results in the multiple top strands of the gate complex. The top strands must be bound to the bottom strand as close as possible to each other. As illustrated in Figure 6, the multiple loop is designed in such a way that after the top hairpin is removed, the dangling end $2a$ preferably binds to $2a^*$ and extends the stem $2b$ to generate recognition domain 2. Another dangling end must be inert in order not to cause any problem by interacting with other molecules.

The recognition sites of enzymes do not remain on the resulting gate complex. This fact keeps the number of required enzymes constant regardless of how many different gate complexes are to be designed, since we are free to choose the design of the DNA bases that are removed and are not restricted by the identifier strands of the gates.

It is crucial to subject this portion of the design to experimental validation. In particular, it is not known whether the restriction enzyme can reliably cleave so close to the three-way branched junction as shown in Figure 6. If not then another technique would need to be devised to make the design scalable. However, even if a small but substantial yield of DNA hairpin precursors are successfully cleaved, then one would need to add a single purification step to segregate them from the uncleaved precursors. Such a step would not require preparing each complex separately.

## 2.4 Toeholds, Linkers, and Buffers

### 2.4.1 Preventing Crosstalk between Linkers

Our design requires a bit of care with the toeholds. Briefly, each toehold can be equal for any "high-level species strands" (such as $A$, $B$, $C$ in Figure 4), and for any buffer strands. However, to avoid unnecessary "crosstalk" between buffers and linkers, and between linkers from two different reactions, it is necessary to use different toehold regions for linkers.

Consider the reactions of Section 2.1.1. The first three reactions share a third reactant $0_o$. This implies that the linker strand of Figure 4 corresponding to these reactions shares the first recognition region $c$ representing $0_o$. If the toehold $t_i$ following the recognition region were equal for these linker strands, then one linker strand could attach to the gate complex $g_1$ of another reaction, erroneously displacing the strand representing $0_o$. To prevent this we use different toeholds for each of these reactions that share their third reactant $0_o$. Similarly, to prevent linkers from displacing non-final reactants (first reactants in a bimolecular reaction, and first or second reactants in a termolecular reaction), we must ensure that $t_i \neq t$ for any $i \in \{1, 2, 3\}$.

This applies to bimolecular high-level reactions as well, but it is routine to verify that no more than three different reactions (the worst case being the case described above) share the same final reactant. It is safe to reuse linker toeholds between reactions that differ in their final reactant because the different initial recognition regions on the linkers will prevent the crosstalk problem described above.

### 2.4.2 Preventing Crosstalk between Buffers

Although we have ensured that buffers and linkers will not have crosstalk because they use different toehold sequences, we must ensure that buffers do not suffer undesired crosstalk with each other. A buffer, unlike a linker, when released uncovers a toehold that is intended to serve as an initial binding site for a strand. Therefore it is not as simple as changing the toehold to prevent buffer crosstalk. Also, a buffer1 strand (a buffer released upon the binding of a first reactant in a reaction, either bimolecular or termolecular) is not garbage-collected as a buffer2 strand is (a buffer released upon the binding of a second reactant in a termolecular reaction). Therefore we must ensure that buffer1 strands are always different from buffer2 strands so that exactly the buffer2 strands are garbage-collected.

To see that this undesired crosstalk does not happen, it suffices to inspect the order of reactants in each reaction. No "high-level species" (corresponding to $A$, $B$, etc. in Figures 3 and 4) is a first reactant in one bimolecular or termolecular reaction but a second reactant in another termolecular reactant. Therefore it is critical to maintain the specific order of the reactants we chose for the high-level reactions of Section 2.1.

## 3 Drawbacks of Design and Comparison with Related Work

This section discusses drawbacks of our design, and compares our design with the most closely related work in terms of the advantages described in Section 1 and the drawbacks described in Section 3.1.

### 3.1 Drawbacks

We have identified four important properties achieved by our proposed design: scalability (as defined in [9]), time-responsiveness, digital abstraction, and energy-efficiency. The design is far from perfect, however. The following are important properties that our design lacks:

**energy-efficiency with non-ideal inputs:** If inputs are given non-ideally, for example concentration 0.99 of $0_w$ and 0.01 of $1_w$ to represent the bit 0 on input wire $w$, then our design must continuously burn fuel to maintain the correct output. Other designs that avoid this problem do so at the cost of time-responsiveness. We conjecture (informally) that given any "realistic" set of molecular primitives, time-responsiveness and this strong form of energy efficiency are incompatible, in the absence of additional external control. Intuitively, it is difficult for a time-responsive circuit to "tell the difference" between a constant imperfect input such as 0.99 of $0_w$ and 0.01 of $1_w$, and a legitimate change of wire $w$ from value 0 to value 1, which must necessarily pass through the non-ideal region of input space during the transition. It seems necessary to use external cues such as a clock signal to trigger re-computation of inputs only after they have changed, in order to keep a constant imperfect input state from being "interpreted" as the start of a change between two ideal input states.

**renewability:** This means that the molecules "representing gates" are not consumed. Which molecules are the "gates" is of course a potentially subjective idea. The main purpose of renewability is that any "fuel" molecule (those consumed in a irreversible reactions) should be "generic", i.e., not specific to any particular gate. Ideally, such molecules are not even specific to the circuit, although one could imagine a first step towards making a design renewable is to design fuel $F_n$ that works for all circuits of size $n$, for instance (perhaps making use of the fact that all "identifiers" in such circuits could be assumed to have the same length, order logarithmic in the number of gates). An ideal goal would be to use chemical reactions in which the only net consumption of fuel is that of ATP. In this case biochemical circuits could be delivered to living cells without needing to supply any external fuel.

**true "detection of absence" of input:** We utilize the dual-rail convention partly because the it helps to modularize the gate design to allow gates to be composed easily. However, it also makes our design less robust because it means that we cannot directly detect the absence of an input species; rather, the absence of $b_w$ is "announced" by the simultaneous presence of $\overline{b}_w$. This may be an unreasonable assumption depending on the nature of the source of input for a biomolecular computation; the absence of a certain microRNA marker indicating disease, for instance, may not coincide with the presence of another marker, or may not be perfectly negatively correlated. What is needed is the design of a "true NOT gate" to convert inputs into the dual-rail inputs required by our design. It is not difficult to achieve this in our design at the cost of energy-efficiency. For instance, each input $1_i$ could be wired to a signal restoration gate (or a cascade of them) along with a constant "threshold" concentration of $0_i$ as input to the same gate. Then, by setting $[1_i] = 0$ or $[1_i] = 2[0_i]$, one could control whether the bit of input wire $i$ is 0 or 1, respectively. However, since $[0_i]$ is always positive, in the case of input bit 1, the input is not ideal so some energy will always be consumed at steady-state. In other words it is possible to move the "energy-efficient" checkmark of our design in Table 1 to "detection of absence of input", but we don't know how to check them both (while preserving the other properties).

## 3.2 Comparison with Related Work

We now describe some previous work in implementing Boolean circuits with DNA. This is not an exhaustive review of the literature, but it covers results most closely related to our proposed methods (strand displacement) or goals (such as time-responsiveness).

Goel and Ibrahimi [4] propose a design for DNA circuits based on the basic primitives of restriction enzyme cutting and sticky-end hybridization. Their design is time-responsive but not energy efficient: DNA strands are irreversibly cut and fuel strands consumed continually to maintain steady-state concentrations. They also describe a certain sense in which their design is "scalable" by virtue of the decoupling of design of data, fuel, and gate strands, but their design is not scalable by the Qian/Winfree definition given above. In particular, their design requires the preparation of DNA double-stranded complexes with unbounded sticky ends (length approximately logarithmic in the number of gates in the circuit) on each side of a gate complex, so it is not clear that the preprocessing required to set up the circuit could be prepared in a one-pot reaction (which is not to say it could not be done via an alternative method). Also, their construction appears to requires the design of new restriction enzymes. Finally, although the circuit is described as "digital", by the definition we use, their design is not digital since signals are allowed to degrade at each layer of the circuit, as Figure 3 of [4] demonstrates.

Seelig, Soloveichik, Zhang, and Winfree [10] pioneered the use of toehold-mediated strand displacement for implementing Boolean circuits. Zhang, Turberfield, Yurke, and Winfree [15] simplified and improved the design. Qian and Winfree [9] further simplified the design, introducing a motif they call a *seesaw gate*, constructed from strand-displacement reactions, and showed that their design is scalable, with the definition of scalable motivated in part by the difficulty of experimental implementations of [10,15], which required separate purification of DNA strands to prepare. The designs of [9,10,15] have an implementation of signal restoration, meaning that if 0 is roughly defined as "between 0.0 and 0.3 (low) concentration of some species", and 1 is defined as "between 0.7 and 1.0 (high) concentration of some species", then imprecision in the precise values of "low" and "high" are corrected at each layer of the circuit. However, this is accomplished at the cost of time-responsiveness: in [9], for instance, "low" concentration of an input species is converted in a one-way reaction to 0 concentration, by reacting the input species with a threshold species that is permanently used up.

Finally, our design borrows many motifs from the construction of Soloveichik, Seelig, and Winfree [11,12], who show how to approximate *any* system of abstract unimolecular and bimolecular chemical reactions (a *chemical reaction network*, or *CRN*) using DNA strand displacement reactions. One could ask, why not simply take an existing abstract chemical implementation of Boolean circuits (for instance, [7]) and apply the construction of [11,12]?

- The DNA complexes required in [11,12] are not scalable by the Qian/Winfree definition. Our fix, using multi-hairpin precursors, is not difficult but is necessary.

| property | this | QW | GI | SSWtr | SSWic |
|---|---|---|---|---|---|
| scalable | ✓ | ✓ | | ✓ | ✓ |
| time-responsive | ✓ | | ✓ | ✓ | |
| digital | ✓ | ✓ | | ✓ | ✓ |
| energy-efficient | ✓ | N/A | | | N/A |
| energy-efficient with non-ideal inputs | | N/A | | | N/A |
| renewable | | | ✓ | | |
| detection of absence of input | | | | ✓ | ✓ |

Table 1: Comparison of some molecular Boolean circuit designs. The definitions of the properties are given in Sections 1 and 3.1. Systems that are not time-responsive are trivially energy-efficient even with non-ideal inputs, for the trivial reason that just enough fuel must be supplied to allow the system to compute once, after which the system becomes "energy-efficient" simply because it has no fuel to consume. For this reason we consider energy-efficiency only for time-responsive systems and mark those fields for non-time-responsive systems with N/A. Also, we have marked the last two as "scalable" since the same hairpin precursors technique could be applied to [11, 12] as well. this = present paper; QW = [9]; GI = [4]; SSWtr = [11, 12] applied to time-responsive CRN described in [7]; SSWic = [11,12] applied to input-consuming CRN such as $A+B \to C$ to implement $C = \mathsf{AND}(A,B)$;

- Our other explicit goals of time-responsiveness, energy-efficiency, and digital abstraction are not necessarily met simultaneously by other chemical reaction designs. Some implementations of Boolean circuits consume their inputs and are hence not time-responsive; for instance, using the reaction $A + B \to C$ to compute $C = \mathsf{AND}(A, B)$ and the pair of reactions $A \to C$ and $B \to C$ to compute $C = \mathsf{OR}(A, B)$. Furthermore, such reactions are not digital since there is no signal restoration. Others, such as that proposed in [7], require termolecular or higher-order reactions not handled by [11, 12].[5] The obvious fix to this, which is to approximate the termolecular reaction $A + B + C \to D$ by the pair of bimolecular reactions $A + B \rightleftharpoons AB$ and $AB + C \to D$, requires implementing the first reversible reaction $A + B \rightleftharpoons AB$ as a pair of irreversible reactions, according to the construction given in [11, 12]. Our design is careful to allow an irreversible reaction only when that reaction moves the system closer to the steady state. Therefore, rather than attempt to carefully design abstract chemical reactions that, when blindly converted by the construction of [11,12], will have no adverse irreversible reactions, we instead chose simply to "hard-code" the desired DNA strand displacement reactions to achieve our design goals. The motifs we used are influenced by [11, 12], but the differences we introduce are nontrivial and are ultimately required to achieve our goal of energy-efficiency. In particular, the DNA complexes shown in Figure 4, required considerable care to design. This is because the second buffer species, unlike the first, must be garbage-collected or it will slow down and, in the presence of non-ideal inputs, ultimately reverse the correct operation of the system.

Table 1 compares our design and those of [4, 9, 11, 12] according the properties discussed in

---

[5]There is also some detail left out of [7] that makes it difficult to analyze the design directly and determine for sure what properties it has; for instance, the actual reactions implementing logic gates are not specified. Table 1 represents our interpretation of the intended design of [7] but it may be mistaken.

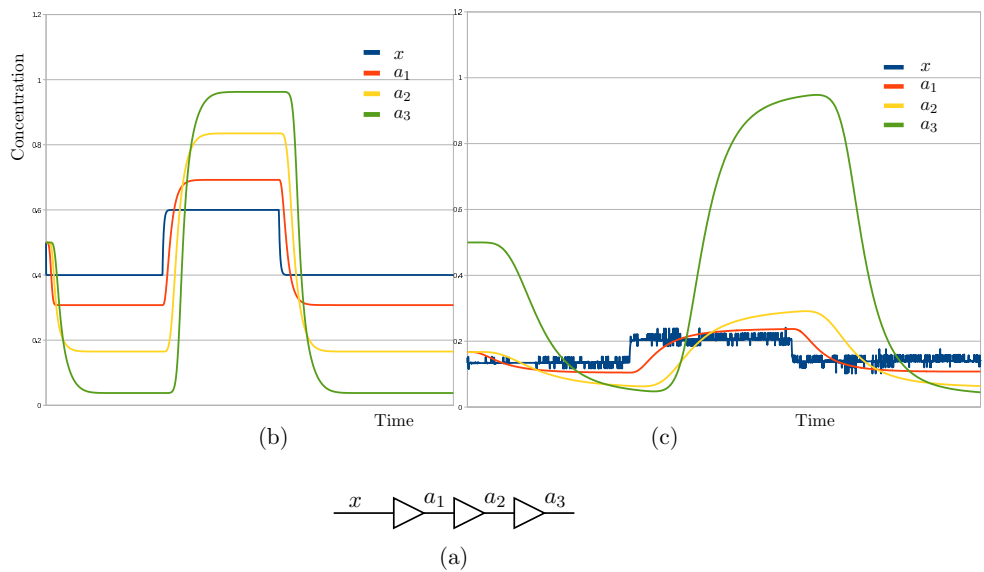

Figure 7: Amplifying a signal three times. Here, high concentration is represented by 1 and low concentration is represented by 0.

Sections 1 and 3.1.

## 4 Simulation Results

We provide a number of examples to illustrate the behavior of our method on some circuits. We do not specify time and concentration units in our simulation charts, as they depend on actual reaction rate constants. Although we do not take these constants into account, they will not change the asymptotic behavior of the system. Therefore, we assume equal reaction rate constants in the simulation. To speed up the simulation, we do not necessarily employ amplifiers on all the wires in the circuit. In the provided schematic diagrams, we show amplifiers wherever they are used. In each example we have used "just enough" amplifiers to justify that the desired output is produced; more amplifiers could move this output closer to ideal in each case.

Figure 7 shows the simulation results of amplifying a signal three times.The concentration of the input $x$ oscillates between 0.4 and 0.6. Figure 7b illustrates the simulation results when abstract chemical reactions are used. Since we use the dual-rail convention, we have two species representing each wire; in all the diagrams, we show only the species representing 1 on a wire. It can be seen that after amplifying $x$ three times, the output concentration gets very close to ideal values, i.e. 0 and 1.

Figure 7c shows the simulation of the underlying strand displacement reactions that implement the abstract high-level reactions of Figure 7b. Note that the values of $x$, $a_1$, and $a_2$ are lower in 7c than what one would expect (7b). This is because of the "buffering effect" discussed in Section 2.2 (also mentioned in [12]): since $x$ (similarly, $a_1$ and $a_2$) is an input to a gate, a portion of molecules representing $x$ quickly equilibrate with gate molecules in which $x$ is a reactant, resulting in a lower concentration of "free" molecules representing $x$. However, this does not affect the behavior of the circuit as long as we ensure each wire is an input to at most one gate. This is the function of the fan-out gates.

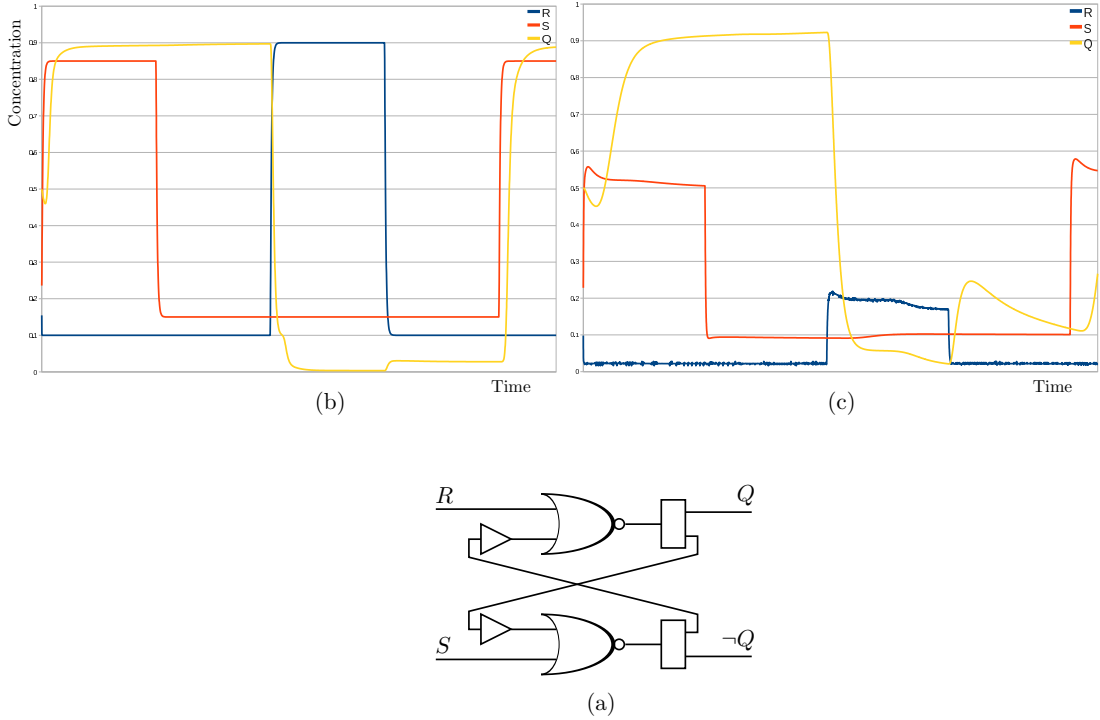

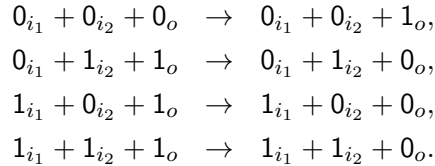

Figure 8: A latch implementation using DNA strand displacement. (a) The schematic diagram. (b) Simulation results using abstract chemical reactions as in Section 2.1. (c) Simulation results using DNA-level reactions as in Section 2.2.

A more complicated example is illustrated in Figure 8. A latch (8a) is a circuit that can store a single bit of information. As can be seen in 8b and 8c, when both inputs $R$ and $S$ have low concentrations (representing 0), the output, $Q$, keeps the state in which it was in "before". Although, in the construction of the latch circuit, it is possible to build the NOR gates using only NAND and fan-out gates (and, possibly, amplifiers), we implemented the NOR gates directly by the following set of reactions:

$$\begin{aligned} 0_{i_1} + 0_{i_2} + 0_o &\to 0_{i_1} + 0_{i_2} + 1_o, \\ 0_{i_1} + 1_{i_2} + 1_o &\to 0_{i_1} + 1_{i_2} + 0_o, \\ 1_{i_1} + 0_{i_2} + 1_o &\to 1_{i_1} + 0_{i_2} + 0_o, \\ 1_{i_1} + 1_{i_2} + 1_o &\to 1_{i_1} + 1_{i_2} + 0_o. \end{aligned}$$

Figure 9 illustrates our simulations on several other circuits.

**Acknowledgment.** The authors are grateful to Erik Winfree for discussing many issues with us.

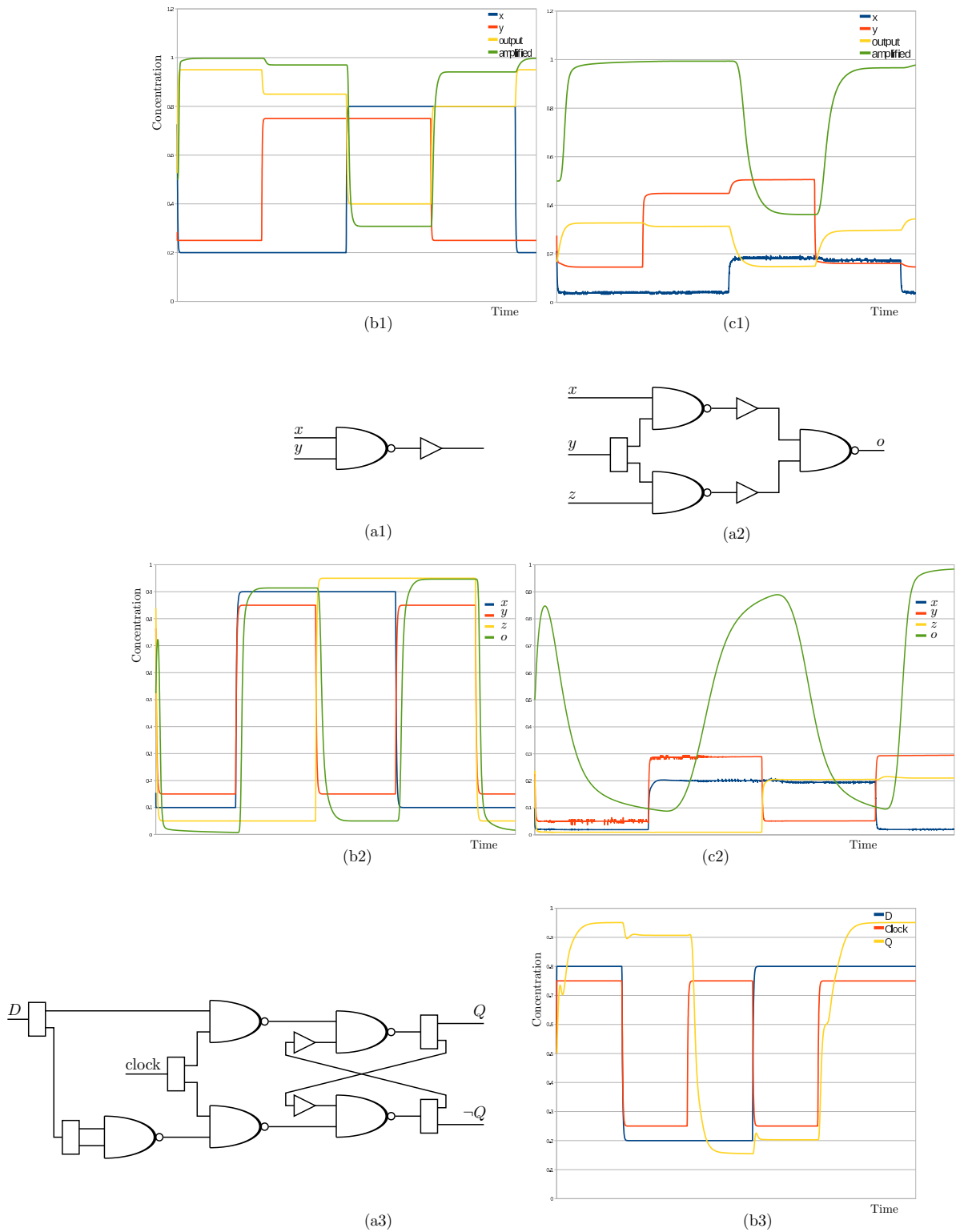

Figure 9: Simulation results using abstract chemical reactions (b1, b2, and b3) and DNA-level reactions (c1 and c2). The schematic diagram in (a3) is a D flip-flop: when the clock is 1, $D$ is copied into $Q$; otherwise, $Q$ is not changed.

# References

[1] Justin S. Bois. *Analysis of Interacting Nucleic Acids in Dilute Solutions*. PhD thesis, California Institute of Technology, 2007.

[2] Robert M. Dirks. *Analysis, Design, and Construction of Nucleic Acid Devices*. PhD thesis, California Institute of Technology, 2005.

[3] *Integrated DNA Technologies*. http://www.idtdna.com.

[4] Ashish Goel and Morteza Ibrahimi. Renewable, time-responsive DNA logic gates for scalable digital circuits. In *DNA15*, pages 67–77, 2009.

[5] Masami Hagiya, Satsuki Yaegashi, and Keiichiro Takahashi. Computing with hairpins and secondary structures of DNA. In *Nanotechnology: Science and Computation*, pages 293–308, 2006.

[6] Joanne Macdonald, Yang Li, Marko Sutovic, Harvey Lederman, Kiran Pendri, Wanhong Lu, Benjamin L. Andrews, Darko Stefanovic, and Milan N. Stojanovic. Medium scale integration of molecular logic gates in an automaton. *Nano Letters*, 6:2598–2603, 2006.

[7] Marcelo O. Magnasco. Chemical kinetics is Turing universal. *Physical Review Letters*, 78(6):1190–1193, Feb 1997.

[8] Robert Penchovsky and Ronald R. Breaker. Computational design and experimental validation of oligonucleotide-sensing allosteric ribozymes. *Nature*, 23:1424–1433, October 2005.

[9] Lulu Qian and Erik Winfree. A simple DNA gate motif for synthesizing large-scale circuits. In *DNA14*, pages 70–89, 2008.

[10] Georg Seelig, David Soloveichik, David Yu Zhang, and Erik Winfree. Enzyme-free nucleic acid logic circuits. *Science*, 314(5805):1585–1588, 2006.

[11] David Soloveichik, Georg Seelig, and Erik Winfree. DNA as a universal substrate for chemical kinetics. In *DNA14*, pages 57–69, 2008.

[12] David Soloveichik, Georg Seelig, and Erik Winfree. DNA as a universal substrate for chemical kinetics. *Proceedings of the National Academy of Sciences*, March 2010.

[13] Milan N. Stojanovic, Tiffany E. Mitchell, and Darko Stefanovic. Deoxyribozyme-based logic gates. *Journal of the American Chemical Society*, 124:3555–3561, April 2002.

[14] Bernard Yurke and Allen P. Mills Jr. Using DNA to power nanostructures. *Genetic Programming and Evolvable Machines*, 4(2):111–122, 2003.

[15] David Yu Zhang, Andrew J. Turberfield, Bernard Yurke, and Erik Winfree. Engineering entropy-driven reactions and networks catalyzed by DNA. *Science*, 318(5853):1121–1125, 2007.